\documentclass[prd, aps, twocolumn, showpacs, preprintnumbers, amsmath, amssymb, nofootinbib, superscriptaddress,10pt]{revtex4-1}
\usepackage[dvips]{graphicx}
\usepackage{graphicx}
\usepackage{epsf}
\usepackage{amsmath}
\usepackage{amssymb}

\usepackage{graphicx}
\usepackage{dcolumn}
\usepackage{bm}
\def\be{\begin{equation}}
\def\ee{\end{equation}}
\def\bea{\begin{eqnarray}}
\def\eea{\end{eqnarray}}

\usepackage{color}

\begin{document}

\title{Asymptotically safe gravity as a scalar-tensor theory \\ and its cosmological implications}

\author{Yi-Fu Cai}
\email{ycai21@asu.edu}
\affiliation{Department of Physics, Arizona State University, Tempe, AZ 85287, USA}
\author{Damien A. Easson}
\email{easson@asu.edu}
\affiliation{Department of Physics, Arizona State University, Tempe, AZ 85287, USA}


\begin{abstract}
We study asymptotically safe gravity with Einstein-Hilbert truncation taking into account the renormalization group running of both gravitational and cosmological constants. We show the classical behavior of the theory is equivalent to a specific class of Jordan-Brans-Dicke theories with vanishing Brans-Dicke parameter, and potential determined by the renormalization group equation. The theory may be reformulated as an $f(R)$ theory. In the simplest cosmological scenario, we find large--field inflationary solutions near the Planck scale where the effective field theory description breaks down. Finally, we discuss the implications of a running gravitational constant to background dynamics via cosmological perturbation theory. We show that compatibility with General Relativity requires contributions from the running gravitational constant to the stress energy tensor to be taken into account in the perturbation analysis.
\end{abstract}

\maketitle

\newcommand{\eq}[2]{\begin{equation}\label{#1}{#2}\end{equation}}

\section{Introduction}

One of the most challenging tasks facing theoretical physicists today is the construction of a consistent ultraviolet (UV) complete theory of gravity.  Weinberg has made the intriguing proposal that the effective description of the quantum nature of a gravitational theory may be non-perturbatively renormalizable via the notion of asymptotic safety (AS) \cite{Weinberg:1977, Weinberg:1979}. In this scenario the renormalization group (RG) flows approach a fixed point in the UV limit and a finite dimensional critical surface of trajectories evolves to this point at short distance scales \cite{Weinberg:2009ca, Weinberg:2009bg, Weinberg:2009wa}. This picture suggests a non-perturbative UV completion for gravity, where the metric fields remain the fundamental degrees of freedom. Moreover, the low energy regime of classical General Relativity is linked with the high energy regime by a well-defined, finite, RG trajectory.

The scenario of AS gravity has been studied extensively in the literature \cite{Kawai:1993mb, Reuter:1996cp, Souma:1999at, Lauscher:2001ya, Lauscher:2001rz, Litim:2003vp, Niedermaier:2006ns, Codello:2007bd, Codello:2008vh, Benedetti:2009rx}. There is evidence that black hole solutions in an AS gravity with Einstein-Hilbert truncation may be nonsingular \cite{Bonanno:2000ep, Bonanno:2006eu, Falls:2010he}; however, the study of black hole physics in an AS gravity theory including higher derivative terms \cite{Cai:2010zh},  showed that while the metric factor may be everywhere finite, curvature invariants may still diverge at the origin. The implication of AS gravity with Einstein truncation and Friedmann-Robertson-Walker (FRW) cosmologies were analyzed with respect to late time cosmological acceleration in Refs. \cite{Bonanno:2001hi, Bonanno:2001xi, Reuter:2004nx, Koch:2007yt, Koch:2010nn, Hindmarsh:2011hx}. The relation between Brans-Dicke theory and AS gravity was discussed in \cite{Reuter:2003ca}. It is also possible that AS gravity might drive inflation at early times \cite{Weinberg:2009wa, Bonanno:2010bt, Tye:2010an, Bonanno:2010mk}.

The majority of work on the subject assumes that the additional contribution to the stress energy tensor due to the running of the gravitational constant $G$ is negligible. It was observed that this additional contribution may play an important role in the background evolution of the early universe \cite{Reuter:2004nx, Koch:2007yt, Koch:2010nn}. In this paper, we study the classical dynamics taking into account the effective stress energy tenor induced by a running $G$. We show that the classical (on-shell) behavior of AS gravity is equivalent to a specific class of Jordan-Brans-Dicke (JBD) \cite{Jordan:1959eg, Brans:1961sx} theories, or an $f(R)$ theory. Through the well-known procedure of making a conformal transformation on the metric, we demonstrate that the theory of AS gravity can be classically reformulated as pure Einstein gravity minimally coupled to a scalar field. Naturally, it is easier to study the cosmological implication of AS gravity in this more familiar Einstein frame. As an application, we consider a toy model with explicit forms for the running gravitational and cosmological constants. In this case, we find that the potential for the scalar is sufficiently flat at large field values to produce slow-roll inflationary solutions. Following the analysis of cosmological perturbations in a generalized JBD model~\cite{Hwang:1995bv, Hwang:1996xh, Tsujikawa:2004my}, a nearly scale-invariant primordial power spectrum can be obtained. However, in this model, the typical energy scale for AS inflation is too high to be compatible with observation. Furthermore, in the classical Einstein frame of the JBD theory, the inflaton potential is not bounded from below and thus the inflaton could fall down to infinity in the future. This pathology is not present in the AS gravity frame due to the natural requirement that the cutoff scale is positively defined. This inconsistency may imply that the toy AS scenario for quantum Einstein gravity is too simple to describe the physics of the RG running over the entire range of scales from the UV to the IR.

Finally, we comment on the significance of the contribution of a running gravitational constant to the stress energy tensor in the AS gravity. In particular, we study the evolution of cosmological perturbations by neglecting the contribution of a running gravitational constant. We find that the sound speed parameter of the gravitational potential is equivalent to that of normal radiation, and is incapable of recovering  standard cosmological perturbation theory in the infrared (IR) limit. We interpret this inconsistency as an indication that it may be inappropriate to neglect the contribution of the running gravitational constant in the study of the perturbation dynamics.

The remainder of this paper is organized as follows. In Section II, we review the idea of AS gravity and describe the basic features of dimensionless coupling parameters in the theory. In Section III, we study the equations of motion in the AS gravity frame and the corresponding JBD gravity frame. At the classical level, we show the two theories are equivalent for a specific choice of parameter and potential for the JBD gravity. Section IV is devoted to cosmological implications of this correspondence. We show how the JBD theory maps to an $f(R)$ theory. We demonstrate the occurrence of inflation in a toy model of AS gravity. In Section V, we study the cosmological perturbation theory of the AS model, ignoring the contribution of running gravitational constant to the background dynamics. We show that ignoring the contribution is inconsistent with General Relativity in the IR limit. Section VI presents a summary and discussion.

We will work with the reduced Planck mass, $M_{pl} = 1/\sqrt{8\pi G_N}$, where $G_N$ is the gravitational constant in the IR limit, and adopt the mostly-plus metric sign convention $(-,+,+,+)$.

\section{Asymptotically safe gravity}

A generally covariant gravitational theory with effective action involving a momentum cutoff $p$ can be expressed as,
\begin{equation}\label{action}
 S_{p}[g_{\mu\nu}]
 = \int{d}^4x\sqrt{-g} \bigg[ p^4g_0(p) +p^2g_1({p})R + \mathcal{O} (R^2) + \cdots \bigg]~,
\end{equation}
where $g$ is the determinant of the metric tensor $g_{\mu\nu}$ and $R$ is the Ricci scalar. The coefficients $g_i$ ($i = 0,\,1,\,\dots$) are dimensionless coupling parameters and are functions of the dimension-full, UV cutoff. In particular, we have
\begin{eqnarray}\label{GLambda}
 g_0 (p) = -\frac{\Lambda(p)}{8 \pi G (p)} \, p^{-4} \,, \qquad
 g_1(p) = \frac{1}{8\pi G(p)}\, p^{-2}\,,
\end{eqnarray}
where $G(p)$ and $\Lambda(p)$ are the quantum corrected gravitational and cosmological constants. The couplings satisfy the following RG equations,
\begin{eqnarray}\label{RGeq}
 \frac{d}{d\ln{p}} \, g_i(p)=\beta_i[g(p)]~.
\end{eqnarray}

According to Ref. \cite{Weinberg:2009ca}, all beta functions vanish when the coupling parameters $g_i$ approach a fixed point $g_{i}^*$ in the scenario of asymptotical safety.
%
%
If $g_{i}^*=0$, the fixed point is Gaussian; if $g_{i}^*\neq0$, the fixed point is Non-Gaussian (NG).  For the NG fixed point, all the coupling parameters are fixed, the cutoff $p$ becomes irrelevant as $p\rightarrow \infty$, and the theory is adequately described by a finite number of higher order counter-terms included in the effective action. Near the fixed point we may Taylor expand the beta functions in a matrix form
\begin{eqnarray}\label{beta}
 \beta_i[g] = \sum_j {\cal B}_{ij}(g_j-g_j^*)~~,
\end{eqnarray}
where the elements of the matrix are defined by ${\cal B}_{ij} \equiv \frac{\partial\beta_i[g]}{\partial{g}_j}^*$ at the fixed point. Solving the RG equations (\ref{RGeq}) in the neighborhood of the fixed point we find
\begin{eqnarray}\label{gi}
 g_i(p)=g_i^*+\sum_m e_i^n \left(\frac{p}{M_*} \right)^{v_n}~,
\end{eqnarray}
where $e^n$ and $v_n$ are the suitably normalized eigenvectors and corresponding eigenvalues of the matrix ${\cal B}_{ij}$. Since ${\cal B}$ is a general real matrix with symmetry determined by a particular gravity model, its eigenvalues can be either real or in pairs of complex conjugates. As a consequence, the dimensionality of the UV critical surface is equal to the number of eigenvalues of the matrix ${\cal B}$, of which the real parts take negative values. The above solution involves an arbitrary mass scale $M_*$. By requiring the largest eigenvector of order unity, $M_*$ is typically identified with the energy scale at which the coupling parameters are just beginning to approach the fixed point.

\section{Dynamics of AS gravity and its classical correspondence}

In this section we analyze the dynamics of AS gravity and show its classical correspondence with scalar-tensor JBD theory.

\subsection{Equations of motion for AS gravity}

Our starting point is a system described by the AS gravity with Einstein-Hilbert truncation, minimally coupled to matter fields,
\begin{eqnarray}
 S_{ASG} = \int d^4x \sqrt{-g} \bigg[\frac{R-2\Lambda}{16\pi G}+{\cal L}_m\bigg]~,
\end{eqnarray}
where $G$ and $\Lambda$ are related to the dimensionless coupling parameters through Eq. (\ref{GLambda}). The Lagrangian ${\cal L}_m$ describes all matter components in this gravitational system.

Varying the action with respect to the metric, one can obtain the following Einstein equation with quantum corrections taken into account,
\begin{eqnarray}\label{EoM_ASG}
 R_{\mu\nu}-\frac{R}{2}g_{\mu\nu} + \Lambda g_{\mu\nu} = 8\pi G T^{(m)}_{\mu\nu} \nonumber\\ + \, G(\nabla_{\mu}\nabla_{\nu}-g_{\mu\nu}\Box)G^{-1}~,
\end{eqnarray}
where we have introduced the covariant derivative $\nabla_{\mu}$ and the operator $\Box \equiv g^{\mu\nu}\nabla_{\mu}\nabla_{\nu}$. The $T^{(m)}_{\mu\nu}$ is the stress energy tensor for all matter components. In the above, the first line is the same as the Einstein equation in classical General Relativity, and the second line denotes a quantum correction due to the RG running of the gravitational constant.

Since both $G$ and $\Lambda$ are no longer constants but functions of the cutoff scale $p$ as reviewed in the previous section, it is important to identify the distribution of $p$ in the quantum corrected manifold. This can be achieved by writing down the generalized Bianchi identity of AS gravity. Assuming the conservation of stress energy tensor of matter fields $\nabla_{\mu}T^{\mu}_{\nu}=0$, we can derive the following useful equation,
\begin{eqnarray}
 - \left( R-2\Lambda \right) \frac{G_{,p}}{G} \, \nabla_\mu p = 2 \Lambda_{,p} \nabla_\mu p~,
\end{eqnarray}
where the subscript $_{,p}$ denotes the derivative with respect to $p$. This equation can be further simplified as
\begin{eqnarray}\label{EoM_p}
 -(R-2\Lambda) \frac{G_{,p}}{G} = 2 \Lambda_{,p}
\end{eqnarray}
when $\nabla_\mu p\neq0$.
%

\subsection{Classical equivalence to the JBD theory}

The JBD theory is an example of a scalar-tensor theory in which the gravitational interaction is mediated by a scalar field as well as the tensor field of General Relativity. The gravitational constant is not presumed to be constant but instead is replaced by a scalar field which can vary from place to place and with time. Compared to classical General Relativity, this theory contains an extra dimensionless constant, $\omega$, which is the so-called Brans-Dicke parameter. The JBD theory with scalar degree of freedom $\varphi$ and potential $U(\varphi)$ is described by the action,
\begin{eqnarray}
 S_{JBD} &=& \int d^4x \sqrt{-g} \bigg[ \frac{\varphi^2R-\omega\nabla_\rho\varphi\nabla^\rho\varphi}{16\pi\varphi} \nonumber\\
 & - & U(\varphi) + {\cal L}_m \bigg]~.
\end{eqnarray}
Depending on the choices of $\omega$ and $U(\varphi)$, this theory includes a variety of classical gravitational theories. For example, classical General Relativity is recovered when $\omega=0$, $U=0$ and $\varphi=1/{G_N}$.

Varying with respect to the metric yields the generalized Einstein equation for the JBD theory,
\begin{eqnarray}\label{EoM_JBD}
 R_{\mu\nu}-\frac{R}{2}g_{\mu\nu} &=& \frac{8\pi}{\varphi} (T^{(m)}_{\mu\nu}-Ug_{\mu\nu}) +\frac{1}{\varphi}(\nabla_{\mu}\nabla_{\nu}-g_{\mu\nu}\Box)\varphi
 \nonumber\\
 &&+ \frac{\omega}{\varphi^2}(\nabla_\mu\varphi\nabla_\nu\varphi -\frac{1}{2}g_{\mu\nu}\nabla_\rho\varphi\nabla^\rho\varphi) ~.
\end{eqnarray}
Varying the action with respect to $\varphi$ gives,
\begin{eqnarray}\label{EoM_varphi_JBD}
 R-16\pi U_{,\varphi}+\frac{2\omega}{\varphi}\Box\varphi-\frac{\omega}{\varphi^2}\nabla_\rho\varphi\nabla^\rho\varphi = 0~,
\end{eqnarray}
where the subscript $_{,\varphi}$ represents differentiation with respect to $\varphi$. Combining this equation with the trace of the generalized Einstein equation, one gets the equation of motion for the scalar field $\varphi$,
\begin{eqnarray}\label{EoM_varphi}
 (2\omega+3)\Box\varphi -8\pi T^{(m)} +32\pi U -16\pi\varphi U_{,\varphi}=0~,
\end{eqnarray}
where $T^{(m)}$ is the trace of the stress energy tensor of matter components.

We now find the specific class of JBD theories which are classically equivalent to AS gravity. Comparing the sets of Einstein equations (\ref{EoM_ASG}) and (\ref{EoM_JBD}), we find the two correspond when we identify,
\begin{eqnarray}\label{condition}
 \varphi &=& G^{-1}~,\nonumber\\
 \omega &=& 0~,\nonumber\\
 U(\varphi) &=& \frac{\Lambda(p)}{8\pi G(p)}~.
\end{eqnarray}
We then verify that the equation of motion for the cutoff $p$ and that for the scalar $\varphi$ are consistent with each other. Making use of Eq. (\ref{EoM_p}) and the trace of Eq. (\ref{EoM_ASG}), we  find
\begin{eqnarray}\label{EoM_G}
 \Box G^{-1} = \frac{8\pi}{3} T^{(m)} -\frac{2\Lambda}{3G} -\frac{2\Lambda_{,p}}{3G_{,p}}~.
\end{eqnarray}
Making use of the correspondence (\ref{condition}), we see that Eq. (\ref{EoM_G}) is exactly consistent with the equation of motion for the scalar in the JBD theory (\ref{EoM_varphi}).

\subsection{Einstein frame description}

In the previous subsection, we have derived a JBD description of the AS gravity at classical level. Consequently, one can reformulate the action of AS gravity in the Jordan frame as,
\begin{eqnarray}
 S_{J} = \int d^4x \sqrt{-g} \bigg[\frac{\varphi R}{16\pi}-U(\varphi)+{\cal L}_m\bigg]~,
\label{jframe}
\end{eqnarray}
where $\varphi$ and $U(\varphi)$ are determined by the condition: Eq. (\ref{condition}). In this subsection, we present an equivalent description of the theory in the Einstein frame. First, we make the following conformal transformation,
\begin{eqnarray}
 \tilde{g}_{\mu\nu} = \Omega^2 g_{\mu\nu}~,~~\Omega^2 = G_N\varphi~,
\end{eqnarray}
where $G_N$ is the value of gravitational constant at the IR limit as defined at the end of the introductory section. As a consequence, the action of AS gravity in the Einstein frame is,
\begin{eqnarray}
 S_{E} &=& \int d^4x \sqrt{-\tilde{g}} \bigg[\frac{\tilde{R}}{16\pi G_N} \nonumber \\
 &-& \frac{1}{2}(\tilde\nabla\sigma)^2 - \tilde{U}(\sigma)+{\cal {\tilde L}}_m\bigg]~,
\label{acteframe}
\end{eqnarray}
where we have introduced a canonical scalar field $\sigma$ defined as
\begin{eqnarray}\label{sigma}
 \sigma \equiv -\sqrt{\frac{3}{2}} M_p \ln{\frac{ \varphi}{8\pi M_p^2}}.
\end{eqnarray}
In the above action, the potential for the new scalar field is determined by
\begin{eqnarray}
 \tilde{U}(\sigma) = \frac{U(\varphi)}{G_N^2\varphi^2} = \frac{\Lambda(p)G(p)}{8\pi G_N^2}~,
\end{eqnarray}
where the second equality is obtained from Eq. (\ref{condition}). Moreover, ${\cal\tilde{L}}_m = {\cal L}_m/G_N^2 \varphi^2$ corresponds to the Lagrangian of other matter fields in the Einstein frame.

\section{Cosmological implication and the AS early universe}

The classical JBD correspondence provides a convenient framework to study any  system in AS gravity. The scenario of AS gravity gives us a new way to understand the quantum nature of gravitational theories. However, in order to test the validity of this scenario, it is necessary to study whether AS gravity is capable of accurately describing the cosmological evolution of our universe. Because the quantum corrections to gravitational theories often become significant at a high energy scale, we expect the theory of AS gravity may shed light on the nature of the universe at early times.

\subsection{Generic analysis of an AS universe}

In a flat FRW background with a scale factor in the Einstein frame, the background equations are given by
\begin{eqnarray}
 \tilde{H}^2 &=& \frac{8\pi G_N}{3} \left(\frac{\sigma^{\prime 2}}{2}+\tilde{U} \right)~,\\
  \tilde{H}^\prime &=& -4\pi G_N \sigma^{\prime 2}~,
\end{eqnarray}
where the prime denotes the derivative with respect to the Einstein frame time coordinate $\tilde t$, given by
 \begin{eqnarray}
 d \tilde t^2 = \exp{\left\{ \sqrt{\frac{2}{3}} \frac{\sigma}{M_p} \right\} } \, dt^2~.
 \end{eqnarray}
Here we have neglected the contribution of normal matter components focusing on the background dynamics of a pure AS cosmological system. From the two background equations, it is well-known that inflationary solutions can be found for sufficiently small values of the  slow-roll parameters,
\begin{eqnarray}
 \epsilon_\sigma &\equiv& \frac{1}{16\pi G_N} \left(\frac{\tilde{U}_{,\sigma}}{\tilde{U}} \right)^2~,\label{sr1}\\
 \eta_\sigma &\equiv& \frac{1}{8\pi G_N}\frac{\tilde{U}_{,\sigma\sigma}}{\tilde{U}}~. \label{sr2}
\end{eqnarray}

Therefore, it is critical to identify the cutoff $p$ as a function of the scalar field $\sigma$ in order to determine if inflation is able to take place. The functional gravitational RG equation is based on a momentum cutoff for the propagating degrees of freedom and captures the non-perturbative information about the gravitational theory. The RG equation is of the form \cite{Reuter:1996cp}:
\begin{eqnarray}
 \frac{\partial}{\partial{\ln{p}}}\Gamma_{p}
 = \frac{1}{2}{\rm{Tr}}\bigg({\delta^{(2)}\Gamma_{p}}+{\bf{R}}(p)\bigg)
   \frac{\partial}{\partial{\ln{p}}}{\bf{R}}(p)~,
\end{eqnarray}
where ${\bf R}$ is an appropriately defined momentum cutoff at the scale $p$ and is usually determined by the so-called optimized cutoff process \cite{Litim:2000ci, Litim:2001up}. In the above formula, we suppose the gauge fixing terms have already been included, although they are irrelevant for our present consideration. Additionally, the trace in the RG equation stands for a sum over spacetime indices and a loop integration. Our philosophy then is to effectively integrate out the high momentum fluctuations with momentum larger than the cutoff $p$, and incorporate them via the modified dynamics for the fluctuations having momentum less than $p$.

\subsection{Early universe in the simplest AS model}

In this subsection, we study the simplest scenario in AS gravity. In general, by solving the beta functions for dimensionless coefficients of AS gravity with Einstein-Hilbert truncation, one finds a Gaussian fixed point in the IR limit and a Non-Gaussian fixed point in the UV limit. The approximate results are given by,
\begin{eqnarray}
\label{G_p}
 G(p) &\simeq& \frac{G_N}{1+\xi_G G_Np^2}~,\nonumber\\
 \Lambda(p) &\simeq& \Lambda_{IR}+\xi_\Lambda p^2~,
\end{eqnarray}
where $G_N$ and $\Lambda_{IR}$ are the values of gravitational constant and cosmological constant in the IR limit. The coefficients $\xi_G$ and $\xi_\Lambda$ are determined by the physics near the UV fixed point. This group of approximate solutions to AS gravity mainly capture the nature of AS of the gravitational theory, but we have neglected many details of the RG flows. Due to this incompleteness, we refer to this model as the {\it simplest} AS model. Explicitly, $\xi_G\simeq0.72$ and $\xi_\Lambda\simeq0.22$ in this case. In the following we will solve the equation of motion of the cutoff scale in the early universe but with energy scale safely below the Planck scale. The requirement of being below the Planck scale guarantees that the Einstein-Hilbert truncation is reliable for the covariant gravity action.

We now show that this theory can be reformulated as a $f(R)$ model. From Eqs. (\ref{EoM_varphi_JBD}) and (\ref{G_p}), we find that
\begin{eqnarray}
 R = 2 \left(\Lambda_{IR}-\frac{\xi_\Lambda}{\xi_G G_N} \right) + \frac{4\xi_\Lambda}{\xi_G}\varphi~,
\end{eqnarray}
allowing the vacuum theory of (\ref{jframe}) to be recast as
\begin{equation} \label{actforr}
 S = \int d^4 \! x  \sqrt{-g} \, f(R) \,,
\end{equation}
with
\begin{eqnarray}
 f(R) = \frac{[G_N(R-2\Lambda_{IR})+2Z]^2}{128\pi G_N^2 Z}~,
\end{eqnarray}
where we have introduced $Z= \xi_\Lambda/\xi_G$.

\begin{figure}[htbp]
\includegraphics[scale=0.45]{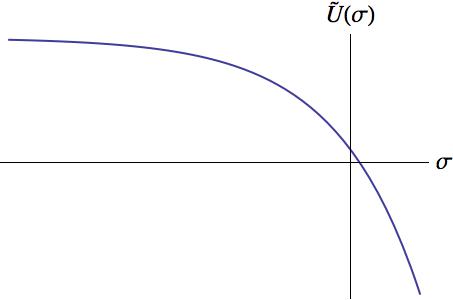}
\caption{Plot of the Einstein-frame potential $\tilde U(\sigma)$.} \label{fig:efpot}
\end{figure}

\begin{figure}[htbp]
\includegraphics[scale=0.45]{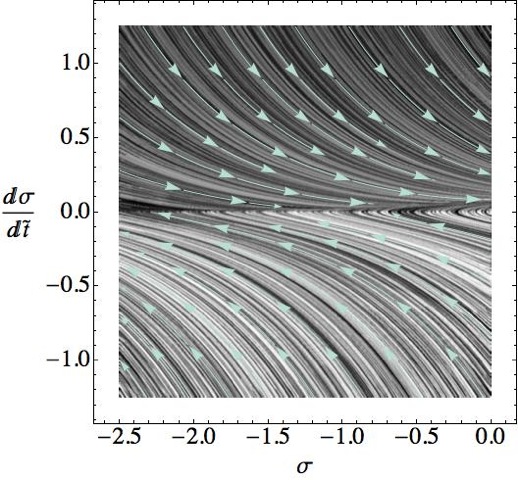}
\caption{Plot of the relevant portion of the $(\sigma, \,\dot \sigma)$--phase space. The inflationary attractor is clearly visible and becomes strong for large values of negative $\sigma$.} \label{fig:phase}
\end{figure}

Moreover, inserting (\ref{G_p}) into the expression (\ref{condition}) and making use of (\ref{sigma}), one gets,
\begin{eqnarray}
 p^2=\frac{e^{-\sqrt{\frac{2}{3}}\frac{\sigma}{M_p}}-1}{\xi_G G_N},
\end{eqnarray}
where we have set $\sigma=0$ at the IR limit. As a consequence, the potential for the inflaton field $\sigma$ is expressed as
\begin{eqnarray}
\label{einpot}
 \tilde{U}(\sigma) = \frac{1}{8\pi G_N^2} \bigg[ \frac{\xi_\Lambda}{\xi_G} + \left(\Lambda_{IR}G_N-\frac{\xi_\Lambda}{\xi_G} \right) e^{\sqrt{\frac{2}{3}}\frac{\sigma}{M_p}} \bigg].
\end{eqnarray}
From the form of the potential, we can find that, when $\sigma\rightarrow-\infty$, $\tilde{U}$ approaches a positive constant. The potential (\ref{einpot}) is plotted in Fig.~\ref{fig:efpot}.  As a consequence, the universe can experience a period of inflationary expansion for sufficiently large values of $\sigma$, when the slow-roll parameters (\ref{sr1}),  (\ref{sr2}), are sufficiently small. However, the energy scale of this scenario is as high as the Planck scale. In this simplest model, the amplitude of quantum fluctuation could catch up with the background evolution and thus we can not trust the effective field description. The portion of the inflationary phase space for large negative $\sigma$ is depicted in Fig.~(\ref{fig:phase}).

Moreover, in the IR limit, we find that the simplest AS scenario fails. When $\sigma \rightarrow 0$, from the equivalent JBD model, we see the scalar field will cross $0$ and then fall down to positive infinity (see Fig.~\ref{fig:efpot}). Of course, this behavior is pathological. In the original AS formulation, the cutoff scale is positively defined and thus the scalar $\sigma$ is expected to be frozen at $0$. This inconsistency implies that, if the covariant gravitational theory has the AS behavior, the RG flows of those constants should not be as rudimentary as the ansatz shown in (\ref{G_p}) in the IR limit.

Note that, in the model we considered, both the metrics of Einstein frame and Jordan frame are inflating. One can define the Hubble parameter in the Jordan frame as
$
 H \equiv \frac{da}{a dt} ~,
$
and its relation to the Hubble parameter in Einstein frame can be expressed as,
$$
 H = e^{\frac{\sigma}{\sqrt{6}M_p}} (\tilde{H}+\frac{\sigma'}{\sqrt{6}M_p}) ~.
$$
Since in the Einstein frame the universe is inflating and the scalar field $\sigma$ satisfies the slow-roll condition, we can conclude that $H$ is also nearly constant which implies the metric of Jordan frame $g$ inflates as well.

\section{Cosmological implications of running gravitational constant}

In the framework of AS gravity, there is one issue which has generated significant debate in the literature. That is, whether running $G$ and $\Lambda$ should bring additional contributions to the stress energy tensor as shown in the last term of Eq. (\ref{EoM_ASG}) \cite{Reuter:2004nx, Koch:2007yt, Koch:2010nn} or not \cite{Bonanno:2001hi, Bonanno:2001xi, Hindmarsh:2011hx}. This question is related to how we connect the AS scenario well studied in a strong gravity system with ordinary Einstein gravity in the weak gravity limit.
%
%
Based on the simple background analysis, it is unclear how to address this issue. Here we propose to answer this question in the context of cosmological perturbation theory.

Our starting point is a universe described by AS gravity with Einstein-Hilbert truncation minimally coupling to matter fields. Without loss of generality, we take a canonical scalar field $\phi$ description of matter with Lagrangian,
\begin{eqnarray}
 {\cal L}_m=-\frac{1}{2}\partial_\mu\phi\partial^\mu\phi-V(\phi)~,
\end{eqnarray}
where $V(\phi)$ is the potential of the background scalar.

\subsection{RG modified background cosmology}

We consider the metric of a homogeneous flat Friedman-Robertson-Walker space-time,
\begin{eqnarray}
 ds^2  \, = \, -dt^2 + a^2(t)d\vec{x}^2~,
\end{eqnarray}
where $t$ is the physical time, ${\vec{x}}$ denotes the co-moving spatial coordinates, and $a(t)$ is the scale factor. By varying the action with respect to the metric, we can obtain the background equations of motion. Both the expansion rate of the universe, i.e., the Hubble parameter, $H\equiv\dot{a}/a$ and its time derivative $\dot H$, are determined by the RG modified Friedmann equations as follows,
\begin{eqnarray}
\label{H2}
 H^2 &=& \frac{8\pi G(p)}{3} \left[ \rho_m+\rho_\Lambda(p) \right]~,\\
\label{dotH}
 \dot{H} &=& -4\pi G(p) (\rho_m+P_m)~.
\end{eqnarray}

In the above, $\rho_m$ and $P_m$ are the energy density and pressure of the matter field $\phi$ respectively:
\begin{eqnarray}
 \rho_m = \frac{1}{2}\dot\phi^2+V~,~~P_m = \frac{1}{2}\dot\phi^2-V~.
\end{eqnarray}
The dynamics of $\phi$ are determined by the equation of motion,
\begin{eqnarray}\label{EoM_phi}
 \ddot\phi+3H\dot\phi+V_{,\phi}=0~,
\end{eqnarray}
where the subscript ``$_{,\phi}$" denotes the derivative with respect to $\phi$. Making use of Eq. (\ref{EoM_phi}), the continuity equation for the matter field is satisfied automatically.

Additionally, we have introduced an effective energy density of the cosmological constant
\begin{eqnarray}
 \rho_\Lambda = \frac{\Lambda(p)}{8\pi G(p)}~,
\end{eqnarray}
which is varying along with the running of the cutoff scale $p$. Therefore, we need to know the dynamics of the cutoff in order to determine the effective energy density of cosmological constant. The running of the cutoff is governed by the generalized Bianchi Identity under the RG corrections, which is given by
\begin{eqnarray}\label{EoM_pp}
 \left[ (3H^2-\Lambda)G_{,p} + G\Lambda_{,p} \right]\dot{p}=0~.
\end{eqnarray}
From Eq. (\ref{EoM_pp}), one can easily find that there exist two solutions for the cutoff scale $p$. One is $\dot{p}=0$ which implies $p$ does not change during the cosmological evolution, but this solution does not help us to understand the role of quantum gravity in the context of cosmology. The other solution is required to satisfy the following equation,
\begin{eqnarray}\label{EoM_p2}
 H^2=\frac{\Lambda}{3}-\frac{G\Lambda_{,p}}{3G_{,p}}~.
\end{eqnarray}

\subsection{RG modified cosmological perturbation theory}

As a consequence, using Eqs. (\ref{H2}), (\ref{dotH}), (\ref{EoM_phi}), and (\ref{EoM_p2}), we can study the cosmological evolution of the universe once suitable initial conditions are chosen. Relevant topics were discussed in the literature; however, in the current paper, our main purpose is to extend the AS cosmology to perturbative level. In particular, we will explore the cosmological perturbation theory of the RG modified universe.

We conduct our analysis in the longitudinal gauge which only involves scalar-type metric fluctuations: \footnote{We refer the reader to \cite{Mukhanov:1990me} for a comprehensive study of cosmological perturbation theory in the classical frame of standard Einstein gravity. }
\begin{eqnarray}
 ds^2 = -(1+2\Phi)dt^2+a^2(t)(1-2\Psi)d\bf{x}^2~,
\end{eqnarray}
thus, as usual, the scalar metric fluctuations are characterized by two functions $\Phi$ and $\Psi$.

By expanding the gravitational equations of motion to linear order, we obtain the $(0,0)$, $(0,i)$, and $(i,i)$ components of perturbed Einstein equations as follows,
\begin{eqnarray}
\label{pert00}
 \frac{\nabla^2}{a^2}\Psi-3H\dot\Psi-3H^2\Phi &=& 4\pi (G\delta\rho + \rho G_{,p}\delta{p})~,\\
\label{pert0i}
 \dot\Psi+H\Phi &=& 4\pi (G\delta{q} + q G_{,p}\delta{p})~,\\
\label{pertii}
 \ddot\Phi+4H\dot\Phi+2\dot{H}\Phi+3H^2\Phi &=& 4\pi (G\delta{P} + P G_{,p}\delta{p})~.
\end{eqnarray}
Moreover, we assume there is no anisotropic stress, leading to $\Phi=\Psi$ due to the vanishing $(i,j)$ component of the perturbed Einstein equations. In the above perturbation equations, $\rho$, $P$ and $q$ are the energy density, pressure and momentum of the whole universe, respectively. Thus, we have $q=0$ at background level. The perturbations of the above quantities are given by,
\begin{eqnarray}
 \delta\rho &=& \dot\phi(\delta\dot\phi-\dot\phi\Phi)+V_{,\phi}\delta\phi+\rho_{\Lambda,p}\delta{p}~,\\
 \delta{P} &=& \dot\phi(\delta\dot\phi-\dot\phi\Phi)-V_{,\phi}\delta\phi-\rho_{\Lambda,p}\delta{p}~,\\
 \delta{q} &=& \dot\phi\delta\phi~,
\end{eqnarray}
respectively. Moreover, the fluctuation of scalar field $\delta\phi$ obeys the perturbed Klein-Gordon equation,
\begin{eqnarray}\label{pert_phi}
 \delta\ddot\phi +3H\delta\dot\phi -\frac{\nabla^2}{a^2}\delta\phi +V_{,\phi\phi}\delta\phi = 4\dot\phi\dot\Phi - 2V_{,\phi}\Phi~.
\end{eqnarray}

Note that, in Eqs. (\ref{pert00}), (\ref{pert0i}) and (\ref{pertii}), the last terms are contributions from the quantum corrections of AS gravity. For the background cosmology, the cutoff scale $p$ is only a function of cosmic time. Thus, the homogeneity of the universe requires that the RG modification to gravity is globally the same in the entire universe. However, after taking into account cosmological perturbations, such a requirement would violate the local gauge transformation of General Relativity. Therefore, we expect the cutoff $p$ ought to be both time and space-dependent when cosmological perturbations are considered. See also \cite{Contillo:2010ju} for an application of space-dependent cutoff scale to cosmological perturbation theory. From the above generic analysis of cosmological perturbations, this feature is reflected in the perturbed mode $\delta{p}(t,\vec{x})$. In the following, we will show that this important correction brought by AS gravity drastically affects the analysis of cosmological perturbation theory.

Before performing the detailed perturbation analysis, we shall study the determination of $\delta{p}$. Recall that the Bianchi Identity $\nabla_\mu G^\mu_\nu$ still holds but its detailed form is modified due to quantum corrections of RG running. At next-to-leading order, it yields,
\begin{eqnarray}\label{pert_Bianchi}
 \frac{d}{dt}\delta(G\rho) &+& 3H\delta(G(\rho+P)) \nonumber \\
 &-& \frac{\nabla^2}{a^2}\delta(Gq) = 3G(\rho+P)\dot\Phi~.
\end{eqnarray}
The combination of (\ref{pert_phi}) and (\ref{pert_Bianchi}) leads to the following interesting relation,
\begin{eqnarray}\label{delta_p}
 \delta{p} = \frac{G_{,p}^2 [\dot\phi(\delta\dot\phi-\dot\phi\Phi) +V_{,\phi}\delta\phi] }{GG_{,pp}\rho_{\Lambda,p} -2G_{,p}^2\rho_{\Lambda,p} -GG_{,p}\rho_{\Lambda,pp}}~,
\end{eqnarray}
when $\dot{p}\neq0$.

Notice that, since the background variable $q$ is vanishing, Eq. (\ref{pert0i}) yields
\begin{eqnarray}\label{delta_phi}
 \delta\phi=\frac{\dot\Psi+H\Phi}{4\pi G\dot\phi}~,
\end{eqnarray}
which is the same as the relation in the classical Einstein gravity. Thus, to make use of the expressions (\ref{delta_p}) and (\ref{delta_phi}), we find there is only a single degree of freedom in the perturbation theory of AS cosmology with Einstein-Hilbert truncation. Combining  Eqs. (\ref{pert00}) and (\ref{pertii}), and making use of Eqs. (\ref{delta_p}) and (\ref{delta_phi}), we obtain the form of main perturbation equation for the gravitational potential,
\begin{eqnarray}
 &&\ddot\Phi +\left[ (3c_s^2-2)H-\frac{\ddot{H}}{\dot{H}}+\frac{G_{,p}}{G}\dot{p} \right] \dot\Phi \nonumber\\
 &&+ \left[2\dot{H}-\frac{H\ddot{H}}{\dot{H}}+3(c_s^2-1)H^2+\frac{H G_{,p}}{G}\dot{p} \right]\Phi-c_s^2\frac{\nabla^2}{a^2}\Phi \nonumber\\
 &&=0~,
\end{eqnarray}
where we have introduced an effective sound speed parameter
\begin{eqnarray}\label{cs2}
 && c_s^2  =  \nonumber \\
 && \frac{G_{,p}^3P - 3GG_{,p}^2\rho_{\Lambda,p} +G^2G_{,pp}\rho_{\Lambda,p} -G^2G_{,p}\rho_{\Lambda,pp}}{G_{,p}^3\rho -GG_{,p}^2\rho_{\Lambda,p} +G^2G_{,pp}\rho_{\Lambda,p} -G^2G_{,p}\rho_{\Lambda,pp}}.
\end{eqnarray}

This main perturbation equation reduces to the standard version in the theory of classical Einstein gravity when we set $\dot{p}=0$ and $c_s^2=1$ by hand. However, we observe that these two perturbation equations are not continuously related by setting a fixed value of cutoff. This feature is completely different from the background dynamics, since once we fix a specific value of $p$ throughout the whole cosmic evolution, the background equations of motion in AS cosmology could automatically become the equations obtained in classical cosmology. Therefore, an important lesson we learned from AS gravity in cosmology is that the dynamics of perturbations might be greatly different from that of the background.

Specifically, we make use of the simplest AS scenario as an example. Inserting (\ref{G_p}) into the expression for the sound speed (\ref{cs2}), one gets
\begin{eqnarray}
 c_s^2 &\simeq& \frac{\xi_\Lambda(1+\xi_GG_Np^2)^2 -8\pi G_N^2\xi_GP_m}{3\xi_\Lambda(1+\xi_GG_Np^2)^2 -8\pi G_N^2\xi_G\rho_m} \nonumber\\
 &\simeq& \frac{1}{3}~,
\end{eqnarray}
if we only focus on the physics below the Planck scale with $p\ll{1}/{\sqrt{G_N}}$. This result implies that the propagation of cosmological perturbations is similar to that of fluctuations from radiation, although we did not introduce any radiation in the model we consider. This result obviously contradicts the usual behavior of cosmological perturbation theory in the frame of standard General Relativity, in which the sound speed of the gravitational potential in a system only filled with gravity and a canonical scalar is unity. Consequently, we arrive at the important result: the contribution of the running gravitational constant to the stress energy tensor should be taken into account in order to maintain consistency of the perturbation analysis with General Relativity. Thus, to determine the perturbation analysis for asymptotically safe cosmology with Einstein-Hilbert truncation, we may simply work directly in the Einstein frame given by (\ref{acteframe}).

\section{Conclusions and Discussion}

The scenario of AS gravity predicts that RG flows of gravitational couplings approach a certain nontrivial fixed point at UV limit. It is important to question how such quantum behavior might be observed in experiments. Unfortunately, this quantum behavior often occurs at an experimentally inaccessible energy scale which is near or even higher than the Planck scale. It becomes almost impossible to test the AS gravity in any laboratory experiments. One might hope that cosmology could provide a new window to observe the quantum nature of Einstein gravity. For example, in the literature, there have been studies of how AS gravity might affect dark energy dynamics, lead to inflationary cosmology by virtue of varying cosmological constant, and be constrained by astronomical observations.

In this paper, we have explored a correspondence between the classical dynamics of AS gravity and a class of JBD theories. In certain cases, the JBD theory can be reformulated as a model of $f(R)$ gravity. Through a conformal transformation, the theory of AS gravity can be expressed as a classical Einstein gravity minimally coupled to a canonical scalar field. As a consequence, one can considerably simplify the analysis of the background dynamics of AS gravity.

As an illustration, we considered a toy model of FRW cosmology with explicit running behavior of both gravitational and cosmological constants. In this simple model, we found that inflation can occur at high Planck scale energies and for sufficiently large values of the inflaton field. Therefore, such a simple model can not describe a realistic universe since the amplitude of quantum fluctuations could be of the same order as the background and would spoil the effective field theory description.

Finally, we studied the significance of the contribution of the running gravitational constant to the stress energy tensor in the frame of AS gravity. In a detailed analysis, we neglected this part of contribution and examined the cosmological perturbation theory. Our results showed that the sound speed parameter in this case coincides with that of normal radiation in the simplest AS model when the cutoff scale is below Planck scale, and one can never recover the standard result in the IR limit. This inconsistency illustrates that the treatment of neglecting the contribution of a running gravitational constant, which may be approximately accurate to describe the background evolution, is not adequate in the study of cosmological perturbations.
Notice that, the correct value of sound speed parameter for metric perturbation can be recovered after we take into account the contribution of the running gravitational constant to the stress energy tensor in AS gravity. In that case, the sound speed equals to unity if the background scalar field is canonical.

\begin{acknowledgments}
The work of Y.F.C. and D.A.E is supported in part by the Cosmology Initiative at Arizona State University. Y.F.C. wish to thank Professor Xinmin Zhang and the Theory Division of the Institute for High Energy Physics for hospitality during the period when the draft of this paper was initiated.
\end{acknowledgments}

\end{document}